\documentclass[11pt,thmsa]{article}
\usepackage{amssymb}
\usepackage{graphicx}

\input{tcilatex}
\begin{document}

\begin{titlepage}
\begin{flushright}
Berlin Sfb288 Preprint \\
physics/9911046
\end{flushright}
\vspace{0.5cm}
\begin{center}
{\Large {\bf   Existence criteria for stabilization

 from the scaling behaviour of ionization probabilities} }

\vspace{1.8cm}
{\large C. Figueira de Morisson Faria,$^{\dag}$  A. Fring$^{*}$ 
and R. Schrader$^{*}$     }\footnote{e-mail addresses: Faria@mpipks-dresden.mpg.de,

Fring@physik.fu-berlin.de,

Schrader@physik.fu-berlin.de}

\vspace{0.5cm}
{${\dag}$ \em  Max-Planck-Institut  f\"{u}r Physik komplexer Systeme, \\
 N\"{o}thnitzer Str. 38, D-01187 Dresden,  Germany\\
 $*$ Institut f\"ur Theoretische Physik,\\
Freie Universit\"at Berlin, Arnimallee 14, D-14195 Berlin, Germany\\}

\end{center}
\vspace{1.2cm}

\renewcommand{\thefootnote}{\arabic{footnote}}
\setcounter{footnote}{0}

\begin{abstract}
We provide a  systematic derivation of  the scaling behaviour of various quantities 
and establish in particular  the scale invariance of the ionization probability.  We discuss 
the gauge invariance of  the scaling properties and the manner in which they can be exploited
 as a consistency check in  explicit analytical expressions, in perturbation theory, 
in the Kramers-Henneberger and Floquet
 approximation, in upper and lower bound estimates and fully numerical solutions of the time
 dependent Schr\"{o}dinger equation. The scaling invariance leads to a differential equation
  which has to be satisfied by the ionization probability and which yields an alternative 
criterium for the existence of atomic bound state stabilization.
\end{abstract}
\vspace{.3cm}
\centerline{November 1999 }
\end{titlepage}
\newpage 

\section{Introduction}

It is up to now still not possible to carry out computations of ionization
probabilities or ionization rates in the high intensity regime in a totally
satisfactory manner. In particular analytical results are extremely rare.
Especially concerning the issue of so-called atomic stabilization \cite
{Gavrila}, numerous computations may be found in the literature which are
contradictory in many cases. Alone for the relatively simple problem of the
one-dimensional delta-potential there exist various recent computations
which do \cite{Su,Geltpro} or do not \cite{Gelt,Mer} support the existence
of stabilization. Roughly speaking, stabilization is the effect that the
ionization probability (or ionization rate to some authors) as a function of
the laser field intensity is decreasing or constant in some region. For
further references and a more detailed terminology, that is, a distinction
into transient, adiabatic, interference or resonance stabilization, see for
instance \cite{Muell}.

It would be highly desirable to settle the controversial issue and pinpoint
possible mistakes, erroneous physical or mathematical assumptions made in
the course of the computations. The main intention of this note is to
contribute to this debate and provide additional alternative consistency
criteria. For this purpose we analyze the scaling behaviour of several
quantities involved in the calculations which address the above-mentioned
problem. This constitutes an adaptation of an idea which has been proved to
be extremely powerful in the context of quantum field theory, for instance
in form of the renormalization group flow (see almost any book on quantum
field theory). In the context of atomic physics, scaling laws have been
considered before \cite{Lamb} in a ``semiempirical'' fashion, as the authors
refer themselves to their own analysis. In order to overcome the slightly ad
hoc way of arguing we provide in this note a systematic derivation of
various scaling laws, which are compatible with the ones proposed in \cite
{Lamb}. In particular, we establish the scale invariance property of the
ionization probability. As a consequence one may exploit these symmetry
properties and check various analytical as well as numerical expressions for
the ionization probability for consistency. In addition, when considering
the ionization probability as a function of various parameters the scale
invariance property allows to trade one particular variable for others. This
permits to interpret and rigorously explain various types of behaviour which
occurred before for more specific situations in the literature. For
instance, for the Hydrogen atom it was found in \cite{PS} that for
increasing principal quantum number the ionization probability decreased and
in \cite{PS2} the opposite behaviour was observed. Our analysis culminates
in the formulation of a simple alternative criterium for the existence of
stabilization.

Our manuscript is organized as follows: In section 2 we derive
systematically the scaling properties of various quantities and establish in
particular the invariance of the ionization probability under scaling. We
show that this property is preserved under gauge invariance. Furthermore,
the scale invariance can be exploited as a consistency check in various
computations. We exhibit this for explicit analytical expressions, for
perturbative calculations, for approximate evaluations in form of
Kramers-Henneberger- and Floquet states and for upper and lower bound
estimates. We demonstrate how the scaling properties can be exploited to
trade some variables for others and use this feature to explain several
types of physical behaviour. As a consequence of our analysis in section 2,
we provide in section 3 a differential equation which has to be satisfied by
the ionization probability and an alternative criterium for the existence of
stabilization. We state our conclusions in section 4.

\section{Scaling behaviour of ionization probabilities}

We consider an atom with potential $V\left( \vec{x}\right) $ in the presence
of a sufficiently intense laser field, such that it may be described in the
non-relativistic regime by the time-dependent Schr\"{o}dinger equation in
the dipole approximation 
\begin{equation}
i\hslash \frac{\partial \psi (\vec{x},t)}{\partial t}=\left( -\frac{\hslash
^{2}}{2m_{e}}\Delta +V\left( \vec{x}\right) +e\,\vec{x}\cdot \vec{E}\left(
t\right) \right) \psi (\vec{x},t)=H\left( \vec{x},t\right) \psi (\vec{x},t).
\label{Schro}
\end{equation}
We further take the pulse to be of the general form 
\begin{equation}
\vec{E}(t)=\vec{E}_{0}f(t)  \label{pulse}
\end{equation}
where $f(t)$ is a function  whose integral over $t$ is assumed to be 
well behaved, with $%
f(t)=0$ unless $0\leq t\leq \tau $. This means $\tau $ is the pulse
duration, $f(t)$ the pulse shape function and $E_{0}$ the amplitude of the
pulse, which we take to be positive  without loss of generality.

Denoting by $\lambda >0$ the dilatation factor and by $\eta $ the scaling
dimension of the eigenfunction $\varphi (\vec{x}):=\psi (\vec{x},t=0)$ of
the Hamiltonian $H\left( \vec{x},t=0\right) $, we consider the following
scale transformations\footnote{%
More formally we could also carry out all our computations by using unitary
dilatation oparators $U(\lambda )$, such that the transformation of the
eigenfunction is described by $U(\lambda )\varphi (\vec{x})=\lambda ^{\eta
}\varphi ^{\prime }(\lambda \vec{x})$ and operators $\mathcal{O}$ acting on $%
\varphi (\vec{x})$ transform as $U(\lambda )\mathcal{O}U(\lambda )^{-1}=%
\mathcal{O}^{\prime }$. } 
\begin{equation}
\vec{x}\rightarrow \vec{x}^{\prime }=\lambda \vec{x}\quad \quad \text{%
and\quad \quad }\varphi (\vec{x})\rightarrow \varphi ^{\prime }(\vec{x}%
^{\prime })=\lambda ^{-\eta }\varphi (\vec{x})\,\,.  \label{dilatation}
\end{equation}
Making the physical assumption that the Hilbert space norm remains
invariant, i.e. $\left\| \varphi (\vec{x})\right\| =\left\| \varphi ^{\prime
}(\vec{x}^{\prime })\right\| $, we deduce immediately that the scaling
dimension has to be $\eta =d/2$, with $d$ being the dimension of the space.
Introducing now the scaling of the dimensional parameters $\hslash ,m_{e}$
and $e$ as 
\begin{equation}
\hslash \rightarrow \hslash ^{\prime }=\lambda ^{\eta _{\hslash }}\hslash
,\quad m_{e}\rightarrow m_{e}^{\prime }=\lambda ^{\eta _{m_{e}}}m_{e}\quad 
\text{and\quad }e\rightarrow e^{\prime }=\lambda ^{\eta _{e}}e\,\,
\label{hme}
\end{equation}
we can scale the whole problem to atomic units, i.e. $\hslash =e=m_{e}$, for
instance by the choice $\lambda =\hslash $, $\eta _{\hslash }=-1$, $\eta
_{e}=-\log _{\hslash }(e)$ and $\eta _{m_{e}}=-\log _{\hslash }(m_{e})$.
Staying for the time being in these units the scaling behaviour (\ref
{dilatation}) may be realized by scaling the coupling constant. Considering
for instance the wavefunction $\varphi (x)=\sqrt{\alpha }\exp (-\alpha |x|)$
of the only bound state when the potential in (\ref{Schro}) is taken to be
the one-dimensional delta-potential $V(x)=\alpha \delta (x)$, equation (\ref
{dilatation}) imposes that the coupling constant has to scale as $\alpha
\rightarrow \alpha ^{\prime }=\lambda ^{-1}\alpha $. Choosing instead the
Coulomb potential in the form $V(\vec{x})=\alpha /r$ requires the same
scaling behaviour of the coupling constant for (\ref{dilatation}) to be
valid. This is exhibited directly by the explicit expressions of the
corresponding wavefunctions $\varphi _{nlm}(\vec{x})\sim \alpha
^{3/2}(\alpha r)^{l}\exp (-\alpha r/n)L_{n+l}^{2l+1}(2\alpha r/n)$ (see e.g. 
\cite{BS}).

From a physical point of view it is natural to require further, that the
scaling behaviour of the wavefunction is the same for all times 
\begin{equation}
\psi (\vec{x},t)\rightarrow \psi ^{\prime }(\vec{x}^{\prime },t^{\prime
})=U^{\prime }(t^{\prime },0)\varphi ^{\prime }(\vec{x}^{\prime })=\lambda
^{-d/2}\psi (\vec{x},t)=\lambda ^{-d/2}U(t,0)\varphi (\vec{x})\,\,.
\end{equation}
Consequently this means that the time evolution operator should be an
invariant quantity under these transformations 
\begin{equation}
U(t_{1},t_{0})=T\left( e^{-\frac{i}{\hslash }\int_{t_{0}}^{t_{1}}H(\vec{x}%
,s)ds}\right) \rightarrow U^{\prime }(t_{1}^{\prime },t_{0}^{\prime
})=T\left( e^{-\frac{i}{\hslash }\int_{\lambda ^{2}t_{0}}^{\lambda
^{2}t_{1}}H^{\prime }(\vec{x},s)ds}\right) =U(t_{1},t_{0})\,\,.
\label{tevol}
\end{equation}
Here $T$ denotes the time ordering. Equation (\ref{tevol}) then suggests
that the scaling of the time has to be compensated by the scaling of the
Hamiltonian and Planck's constant in order to achieve invariance. Scaling
therefore the time as 
\begin{equation}
t\rightarrow t^{\prime }=\lambda ^{\eta _{t}}t\,\,,  \label{tscale}
\end{equation}
with $\eta _{t}$ being unknown for the moment, equation (\ref{tevol}) only
holds if the Stark Hamiltonian of equation (\ref{Schro}) scales as 
\begin{equation}
H\left( \vec{x},t\right) \rightarrow H^{\prime }\left( \vec{x}^{\prime
},t^{\prime }\right) =\lambda ^{\eta _{H}}H\left( \vec{x},t\right) \,\,\quad 
\text{with \quad }\eta _{H}=\eta _{\hslash }\,\,-\eta _{t}\,\,.  \label{HD}
\end{equation}
The properties (\ref{tscale}) and (\ref{HD}) could also be obtained by
demanding the invariance of the Schr\"{o}dinger equation (\ref{Schro}). The
overall scaling behaviour of $H\left( \vec{x},t\right) $ is governed by the
scaling of the Laplacian, the electron mass and Planck's constant, such that
we obtain the further constraint 
\begin{equation}
\eta _{H}=2\eta _{\hslash }-\eta _{m_{e}}-2\,\,.  \label{eh}
\end{equation}
$\,\,$As a consequence we can read off the scaling properties of the
potential as 
\begin{equation}
V\left( \vec{x}\right) \rightarrow V^{\prime }\left( \vec{x}^{\prime
}\right) =\lambda ^{\eta _{H}}V\left( \vec{x}\right) \quad \,\,\,.\,
\label{VE}
\end{equation}
What does this behaviour imply for some concrete potentials? Having scaled
everything to atomic units, relation (\ref{eh}) suggests that $\eta _{H}=-2$%
. Considering for this situation the one-dimensional delta-potential and the
Coulomb potential in the forms specified above, equation (\ref{VE}) imposes
that the coupling constant has to scale as $\alpha \rightarrow \alpha
^{\prime }=\lambda ^{-1}\alpha $ in both cases. This behaviour of the
coupling constant is in agreement with our earlier observations for the
corresponding wavefunctions. We also observe immediately that the behaviour (%
\ref{VE}) may be realized for the general class of Kato small potentials. We
recall that if for each constant $\beta $ with $0<\beta <1$ there exists a
constant $\gamma $, such that $\left\| V\psi \right\| \leq \beta \left\|
-\Delta \psi \right\| +\gamma \left\| \psi \right\| $ holds for all $\psi $
in the domain $\mathcal{D}(-\Delta /2)$, the potential $V$ is called Kato
small. We see immediately that the scaling of the first term is entirely
governed by the Laplacian such that $\beta \rightarrow \beta ^{\prime
}=\beta $ is scale invariant and that $\gamma $ has to scale as $\gamma
\rightarrow \gamma ^{\prime }=\lambda ^{-2}\gamma $ due to the scale
invariance of the norm.

It is intriguing to note that there exists an interesting class of
potentials which scale alone via their dependence on $\vec{r}$ and which do
not contain any energy scale $\alpha $ at all, as for instance $V\left( \vec{%
x}\right) =1/r^{2}$ or the two-dimensional delta potential.

In \cite{Lamb} the interesting proposal was made to exploit the scaling
behaviour in order to use known properties of the Hydrogen atom to
understand the behaviour of Helium. For this purpose the Schr\"{o}dinger
equation describing Helium, i.e. (\ref{Schro}), for the potential $%
V_{He}\left( \vec{x}\right) =-Ze^{2}/r$ and the mass $m_{e}$ replaced by the
reduced mass $\mu $, is scaled to the one which describes Hydrogen.
Translating the quantities of \cite{Lamb} into our conventions this
transformation is realized by $\lambda =(\mu /m_{e})Z$, $\eta _{t}=\log
_{\lambda }(Z^{2}\mu /m_{e})$, $\eta _{\mu }=\log _{\lambda }(m_{e}/\mu )$, $%
\eta _{Z}=-\log _{\lambda }Z$ and $\eta _{\hslash }=\eta _{e}=0$. These
quantities are consistent with the additional constraint $\eta _{H}=2\eta
_{\hslash }-\eta _{m_{e}}-2,$ which results for the potential $V_{He}\left( 
\vec{x}\right) $ from the scaling arguments. We would like to point out that
this is only one of many possible choices. It might be more convenient to
use for instance $\lambda =Z$, $\eta _{t}=2$, $\eta _{\mu }=$ $\eta
_{\hslash }=\eta _{Z}+1=\log _{Z}(m_{e}/\mu )$ and $\eta _{e}=0$ instead.

\noindent We will now consider the constraint resulting from equation (\ref
{HD}) on the scaling behaviour of the pulse. We have 
\begin{equation}
\vec{E}\left( t\right) \rightarrow \vec{E}^{\prime }\left( t^{\prime
}\right) =\lambda ^{\eta _{E}}\vec{E}\left( t\right) \,\,\,\quad \text{%
with\quad }\eta _{E}=\eta _{H}-\eta _{e}-1.\,
\end{equation}
This equation is not quite as restrictive as for the potential, since in the
latter case we could determine the behaviour of the coupling whereas now a
certain ambiguity remains in the sense that we can only deduce 
\begin{equation}
\vec{E}_{0}\rightarrow \vec{E}_{0}^{^{\prime }}=\lambda ^{\eta _{E_{o}}}\vec{%
E}_{0}\,,\quad f\left( t\right) \rightarrow f^{\prime }\left( t^{\prime
}\right) =\lambda ^{\eta _{f}}f\left( t\right) ,\quad \text{with }\eta
_{E_{0}}+\eta _{f}=\eta _{E}\,.
\end{equation}
Thus, under the assumptions we have made, it is not possible to disentangle
the contribution coming from the scaling of the amplitude or the pulse shape
function. However, there might be pulse shape functions for which $h_{f}$
has to be $0$, since no suitable parameter is available in its explicit form
to achieve the scaling.

Finally, we come to the scaling behaviour of the ionization probability.
Denoting by $P$ the orthogonal projection in $L^{2}(I\!\!R^{3})$ onto the
subspace spanned by the bound states of $H\left( \vec{x},t=0\right) $, the
ionization probability turns out to be a scale invariant quantity 
\begin{equation}
\mathcal{P}\left( \varphi \right) =\left\| \left( 1-P\right) U\left( \tau
,0\right) \varphi \right\| ^{2}\rightarrow \mathcal{P}^{\prime }\left(
\varphi ^{\prime }\right) =\mathcal{P}\left( \varphi \right) .  \label{ion}
\end{equation}
This follows by means of (\ref{dilatation}), (\ref{tevol}) and by noting
that the projection operator has to be a scale invariant quantity, i.e. $%
P\rightarrow P^{\prime }=P$. From a physical point of view this is clear
unless we were able to scale bound states into the continuum, which is 
impossible, since negative energies will remain always negative even after
being scaled.
Mathematically this means we have to demand that $P^{\prime }$ and $P$ are
related to each other by a unitary transformation. 

We recapitulate that our sole assumptions in this derivation were to demand
the invariance of the Hilbert space norm, i.e. $\left\| \varphi (\vec{x}%
)\right\| =\left\| \varphi ^{\prime }(\vec{x}^{\prime })\right\| $, and that
the scaling of the wavefunction is preserved for all times.

We shall now utilize this symmetry property in various approaches, which can
be carried out either numerically or analytically. At this point we scale
everything to atomic units which we will use from now onwards.

\subsection{Gauge invariance}

First of all we would like to establish that these scaling properties hold
in every gauge, as one naturally expects. We recall that different gauges
are related by a time-dependent unitary operator $A_{g_{2}\leftarrow
g_{1}}(t)$. For instance the wavefunction in gauge $g_{1}$ and gauge $g_{2}$
are related as $\Psi _{g_{2}}(\vec{x},t)=A_{g_{2}\leftarrow g_{1}}(t)\Psi
_{g_{1}}(\vec{x},t)$. The velocity gauge is obtained from the length gauge
by 
\begin{equation}
A_{v\leftarrow l}(t)=e^{i\vec{b}(t)\cdot \vec{x}}\rightarrow A_{v\leftarrow
l}^{\prime }(t^{\prime })=A_{v\leftarrow l}(t)
\end{equation}
the velocity gauge from the Kramers-Henneberger gauge by 
\begin{equation}
A_{v\leftarrow KH}(t)=e^{-ia(t)}e^{i\vec{c}(t)\cdot \vec{p}}\rightarrow
A_{v\leftarrow KH}^{\prime }(t^{\prime })=A_{v\leftarrow KH}(t)
\end{equation}
and the length gauge from the Kramers-Henneberger gauge by 
\begin{equation}
A_{l\leftarrow KH}(t)=e^{-ia(t)}e^{-i\vec{b}(t)\cdot \vec{x}}e^{i\vec{c}%
(t)\cdot \vec{p}}\rightarrow A_{l\leftarrow KH}^{\prime }(t^{\prime
})=A_{l\leftarrow KH}(t)\,\,.  \label{GKG}
\end{equation}
The defining relations for the classical momentum transfer $\vec{b}(t)$, the
classical displacement $\vec{c}(t)$ and the classical energy transfer $\vec{a%
}(t)$ then yield

\begin{eqnarray}
\vec{b}(t) &=&\vec{E}_{0}b_{0}(t)=\int_{0}^{t}ds\vec{E}(s)\rightarrow \vec{b}%
^{\prime }(t^{\prime })=\int_{0}^{t\lambda ^{2}}ds\lambda ^{-3}\vec{E}%
(s\lambda ^{-2})=\lambda ^{-1}\vec{b}(t)  \label{b1} \\
\vec{c}(t) &=&\vec{E}_{0}c_{0}(t)=\int_{0}^{t}ds\vec{b}(s)\rightarrow \vec{c}%
^{\prime }(t^{\prime })=\int_{0}^{t\lambda ^{2}}ds\lambda ^{-1}\vec{b}%
(s\lambda ^{-2})=\lambda \vec{c}(t)  \label{b2} \\
\vec{a}\left( t\right)  &=&\vec{E}_{0}a_{0}(t)=\frac{1}{2}%
\int_{0}^{t}ds\,b^{2}\left( s\right) \rightarrow \vec{a}^{\prime }(t^{\prime
})=\int_{0}^{t\lambda ^{2}}ds\lambda ^{-2}b^{2}(s\lambda ^{-2})=\vec{a}%
(t)\quad .  \label{b3}
\end{eqnarray}
These quantities scale in the expected manner, that is $\vec{b}(t)$ scales
as a momentum, $\vec{c}(t)$ as the space and $\vec{a}(t)$ remains invariant.
Taking these properties into account, we observe easily that the operator $%
A_{g_2 \leftarrow g_1}(t)$ is an invariant quantity under scaling 
\begin{equation}
A_{g_2\leftarrow g_1}(t)\rightarrow A_{g_2 \leftarrow g_1}^{\prime
}(t)=A_{g_2 \leftarrow g_1}(t)
\end{equation}
for all cases $g_{1}$ and $g_{2}$ mentioned. Hence the scaling behaviour is
preserved in all gauges. It is interesting to note that one may reverse the
logic here and deduce from a broken scale invariance onto a broken gauge
invariance. However, in general  gauge invariance is not broken in such a
crude manner, e.g. in \cite{Mil} (see eqn. (22) therein) the gauge
invariance is broken in a scale invariant fashion.

\subsection{Symmetry properties for analytical expressions of $\mathcal{P}$}

Keeping the pulse shape function invariant under the scaling transformations
we incorporate now the explicit functional dependence into the ionization
probability. The fundamental parameters are the field amplitude, the pulse
length and the coupling constant. The previous observations then suggest
that 
\begin{equation}
\mathcal{P}(E_{0},\tau ,\alpha )=\mathcal{P}(E_{0}^{\prime },\tau ^{\prime
},\alpha ^{\prime })\,\,.  \label{invP}
\end{equation}
Assuming from now on that the coupling constant scales as for the
one-dimensional delta- and the Coulomb potential, the meaning of equation (%
\ref{invP}) is that the ionization probability remains invariant under the
transformations 
\begin{equation}
E_{0}\rightarrow E_{0}^{\prime }=\lambda ^{-3}E_{0},\qquad \tau \rightarrow
\tau ^{\prime }=\lambda ^{2}\tau ,\qquad \alpha \rightarrow \alpha ^{\prime
}=\lambda ^{-1}\alpha \,\,.  \label{tran}
\end{equation}

\noindent We can exploit the symmetry property (\ref{invP}) most easily when
we have an explicit analytical expression for $\mathcal{P}\left( \varphi
\right) $ at hand. Considering for example the $\delta $-potential and
taking the pulse to be the $\delta $-kick, i.e. $E(t)=E_{0}\delta (t),$ $%
b(t)=E_{0}0^{+},$ $c(t)=0,$ the ionization probability of the bound state
was computed to be  \cite{FFS2} 
\begin{equation}
\mathcal{P}\left( \varphi \right) =1-\frac{4}{\pi ^{2}}\left|
\int\limits_{-\infty }^{\infty }dp\frac{\exp \left( -i\tau \alpha ^{2}\frac{%
p^{2}}{2}\right) }{\left( 1+\left( p+b\left( \tau \right) /\alpha \right)
^{2}\right) \left( 1+p^{2}\right) }\right| ^{2}\,\,.  \label{dd}
\end{equation}
Obviously the r.h.s. of (\ref{dd}) passes the test and remains invariant
under the scale transformation in form of (\ref{b1}) and (\ref{tran}).

\subsection{Perturbation theory}

Usually one is not in the fortunate situation to have explicit expressions
for the ionization probability available as in the previous subsection.
However, the symmetry property may also be utilized when computing $\mathcal{%
P}\left( \varphi \right) $ approximately either in a numerical or analytical
fashion. We recall that the essential ingredient of perturbation theory is
to expand the time evolution operator as a series in $E_{0}$ or $\alpha $
for small or large field intensities, respectively. We can formally write

\begin{equation}
U(t_{1},t_{0})=\sum\limits_{n=0}^{\infty }U(n|t_{1},t_{0})\,\,.  \label{se}
\end{equation}
Since $U(t_{1},t_{0})$ is a scale invariant quantity, the invariance
property (\ref{tevol}) must hold order by order, that is for $0\leq n\leq
\infty $ we have 
\begin{equation}
U(n|t_{1},t_{0})\rightarrow U^{\prime }(n|t_{1}^{\prime },t_{0}^{\prime
})=U(n|t_{1},t_{0})\,\,.
\end{equation}
Considering now for instance the high intensity regime and performing
Gordon-Volkov perturbation theory (e.g. \cite{FKS,FFS3}), the first terms in
(\ref{se}) read 
\begin{eqnarray}
U(0|t_{1},t_{0}) &=&\exp (i(t_{1}-t_{0})\Delta /2)=\exp
(i(t_{1}-t_{0})\lambda ^{2}\lambda ^{-2}\Delta /2)=U^{\prime
}(0|t_{1}^{\prime },t_{0}^{\prime })\,  \label{u0} \\
U(1|t_{1},t_{0})
&=&i\int_{t_{0}}^{t_{1}}dsU(0|t_{1},s)VU(0|s,t_{0})=U^{\prime
}(1|t_{1}^{\prime },t_{0}^{\prime })\,\,\,\,.  \label{u1}
\end{eqnarray}
Whilst it was fairly obvious that the general expressions (\ref{u0}) and (%
\ref{u1}) remain invariant under scaling, this consistency check might be
less trivial when carried out after the expressions have been evaluated
explicitly either numerically or analytically.

\subsection{Expansions in terms of Kramers-Henneberger states or Floquet
states}

The essence of the Kramers-Henneberger approximation (e.g. \cite{VPS}) is to
exploit the fact that when the gauge transformation (\ref{GKG}) is carried
out on the Stark Hamiltonian, the term involving the laser pulse disappears
and instead the potential is shifted by the total classical displacement,
i.e. $V(\vec{x})\rightarrow V(\vec{x}-\vec{c}(t))$. Having chosen the pulse
in such a way that $\vec{c}(t)$ is a periodic function in time, one can
expand the shifted potential in a Fourier series 
\begin{equation}
V(\vec{x}-\vec{c}(t))=\sum\limits_{n=-\infty }^{\infty }V_{n}e^{in\omega
t}\,\,.\,  \label{KHA}
\end{equation}
In the Kramers-Henneberger approximation one assumes now that the zero mode
is dominant such that the remaining terms may be treated as perturbations.
From the scaling behaviour of the l.h.s. of (\ref{KHA}) we deduce
immediately that the frequency has to scale inverse to the time, i.e. $%
\omega \rightarrow \omega ^{\prime }=\lambda ^{-2}\omega $, and that each
mode in the series scales in the same way as the potential, i.e. 
\begin{equation}
V_{n}\rightarrow V_{n}^{\prime }=\lambda ^{-2}V_{n}\,\,\,.  \label{VN}
\end{equation}
As an example let us consider the expansion for a square-well potential of
depth $\alpha V_{0}$ and of width $d$ subjected to a pulse of linearly
polarized monochromatic light. The modes are of the general form (first
reference in \cite{VPS}) 
\begin{equation}
V_{n}=|\alpha V_{0}|\,g[(d/2-x)\omega ^{2}/E_{0}]\,\,,  \label{VNS}
\end{equation}
where the explicit formula of the function $g$ is given in term of
Chebyshev-polynomials which, however, is not important for our purpose. We
only need to know that it scales by means of its argument alone. Since the
argument is scale invariant, we observe with the help of (\ref{VE}) for $%
\eta _{H}=2$ in atomic units that (\ref{VN}) holds for each coefficient in (%
\ref{VNS}).

The analysis of the scaling behaviour of the Floquet expansion is very
similar. Instead of exploiting the periodicity of the potential one makes
additional use of the periodicity of the field and expands $\psi (\vec{x}%
,t)=\sum_{n=-\infty }^{\infty }\psi _{n}(\vec{x})e^{in\omega t}$. It is then
obvious by the same reasoning as before that the scaling of the individual
modes has to be the same as for the field itself, i.e. $\psi _{n}(\vec{x}%
)\rightarrow \psi _{n}^{\prime }(\vec{x}^{\prime })=\lambda ^{-d/2}\psi _{n}(%
\vec{x})$.

\subsection{Upper and lower bounds}

In \cite{FKS,FKS2,FFS,FFS3} analytical expressions for upper and lower
bounds, $\mathcal{P}_{u}\left( \varphi \right) $ and $\mathcal{P}_{l}\left(
\varphi \right) $, respectively, were derived and analyzed. Depending on the
particular parameters these expressions put more or less severe constraints
on the actual value of $\mathcal{P}\left( \varphi \right) $, in the sense
that $\mathcal{P}_{l}\left( \varphi \right) \leq \mathcal{P}\left( \varphi
\right) \leq \mathcal{P}_{u}\left( \varphi \right) $. Since $\mathcal{P}%
\left( \varphi \right) $ is a scale invariant quantity, also the bounds have
to respect this symmetry. Otherwise they could be scaled to any desired
value. We present just one example for one particular upper bound (the
arguments carry through equally for lower bounds) to convince ourselves that
this is indeed the case. For instance under the condition $b(\tau
)^{2}/2>-E\equiv $ binding energy of $\varphi $, the following upper bound
was derived in \cite{FKS} 
\begin{equation}
\mathcal{P}_{u}(\varphi )=\Bigg\{ \int\limits_{0}^{\tau }\Vert (V(\vec{x}%
-c(t)e_{z})-V(\vec{x}))\varphi \Vert dt+|c(\tau )|~\Vert p_{z}\varphi \Vert +%
\frac{2|b(\tau )|\Vert p_{z}\varphi \Vert }{2E+b(\tau )^{2}}\Bigg\}^{2}\;.
\label{PU}
\end{equation}
It is easy to see term by term that the r.h.s. of (\ref{PU}) scales
invariantly. In \cite{FFS} we have already exploited this property. In fact,
we found that the bound (\ref{PU}) is only considerably below 1 for very
small values of the pulse length $\tau $. Since the binding energy has to
scale in the same manner as the Hamiltonian $H(\vec{x},t=0)$, that is $%
E\rightarrow E^{\prime }=\lambda ^{-2}E$, we could also, due to the scale
invariance property, enlarge the pulse durations by considering higher
Rydberg states. In this way we could study pulses which are physically more
conceivable, at the cost of having to deal with higher principal quantum
numbers.

\subsection{Trading some variables for others}

Of course the principle mentioned at the end of the last subsection is very
general and we may always trade some variables for others, simply by
bringing the relevant $\lambda $'s in (\ref{tran}) to the other side of the
equation. For instance from $\mathcal{P}(\lambda ^{3}E_{0},\tau ,\alpha )=%
\mathcal{P}(E_{0},\lambda ^{2}\tau ,\lambda ^{-1}\alpha )\,\,$it follows
that instead of varying the field amplitude and keeping $\tau $ and $\alpha $
fixed, we could equivalently keep $E_{0}$ fixed and vary simultaneously $%
\tau $ and $\alpha $ in the described fashion. As a consequence we can give
some alternative physical interpretation to the extreme intensity limit
considered in \cite{FKS2,FFS2} 
\begin{equation}
\lim\limits_{E_{0}\rightarrow \infty }\mathcal{P}(\varphi )=\lim\limits\Sb %
\tau \rightarrow \infty  \\ \alpha \rightarrow 0  \endSb \mathcal{P}(\varphi
)\,\,.
\end{equation}
This means switching off the potential and exposing the atom to an
infinitely long pulse with some finite field amplitude is equivalent to
keeping the pulse length and the coupling constant finite and taking the
field amplitude to infinity.

\vspace*{-0.4cm}

\begin{center}
\includegraphics[width=10cm,height=11.5cm,angle=0]{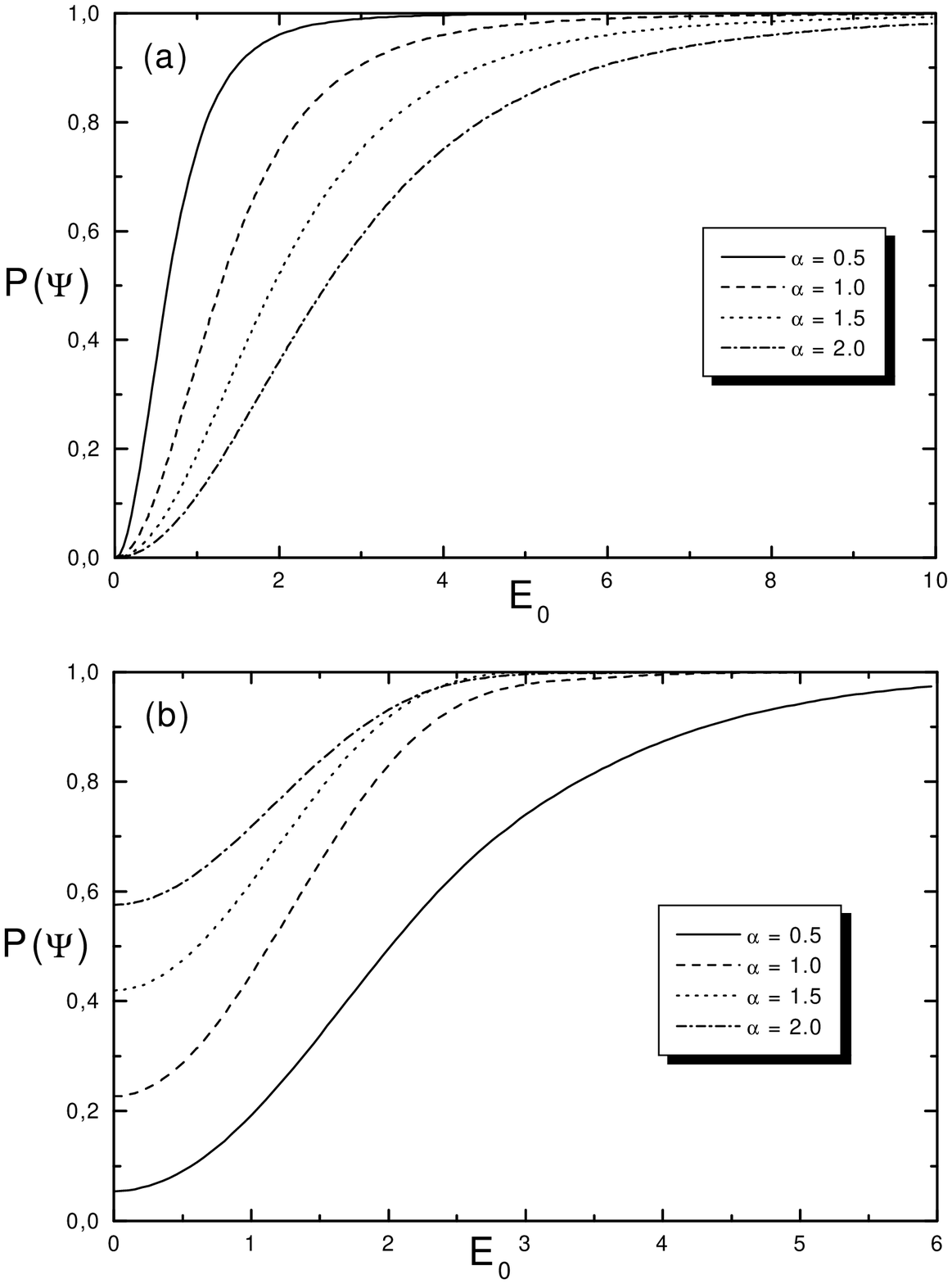}
\end{center}


\noindent {\small Figure 1: Part (a) shows the ionization probability
as a function of the field amplitude $E_{0}$ for a $\delta $-potential atom
subjected to a $\delta $-kick pulse (\ref{dd}) for $\tau =0.001$ and various
coupling constants. Part (b) shows the ionization probability to
zeroth order Gordon-Volkov perturbation theory as a function of the field
amplitude $E_{0}$ for a $\delta $-potential atom subjected to a double $%
\delta $-kick pulse of the form $E(t)=E_{0}(\delta (t)-2$}$\delta (t-\tau ))$
{\small for $\tau =1.1$ and various coupling constants. Notice that for this
pulse the conditions $b(\tau )=0$ and $c(\tau )\neq 0$ hold. For a detailed
derivation see \cite{FFS2}.}


We can also use the scale invariance property to give a simple explanation
to a behaviour, which at first sight appears somewhat puzzling. In \cite
{FFS2,PS} it was observed that the ionization probability is sometimes a
decreasing and sometimes an increasing function of the coupling constant
when the other parameters are kept fixed, refer to figure 1.

Important for the explanation of this feature is that in the former case $%
b(\tau )=0,c(\tau )\neq 0$ and in the latter $b(\tau )\neq 0,c(\tau )=0$.
Assuming now that the dependence of the ionization probability on the field
amplitude enters only through the quantities $b(\tau )$ and $c(\tau )$ and
in addition that the dependence on the pulse length is very weak in
comparison with the one on $b(\tau ),c(\tau )$ and $\alpha $, according to
the scale invariance property we can write 
\begin{equation}
\mathcal{P}(b(\tau ),c(\tau ),\alpha )\approx \mathcal{P}(\lambda
^{-1}b(\tau ),\lambda c(\tau ),\lambda ^{-1}\alpha )\,\,.  \label{bcp}
\end{equation}
Thus, in case the functional dependence on $c(\tau )$ is much
weaker than the one on $b(\tau )$, we have to increase the coupling constant
when the total classical momentum transfer is increased in order to keep the
ionization probability fixed. Noting that $E\sim \alpha ^{-2}$, this is
expected from the classical point of view, since to free a more deeply bound
state with the same probability requires a larger momentum transfer. In the
reversed case, in which the functional dependence on $c(\tau )$ is much
stronger than the one on $b(\tau )$ we have to decrease the coupling
constant when the total classical displacement is increased in order to keep
the ionization probability at the same value. Also this behaviour is
expected from a classical point of view, since when a less deeply bound
state is freed with the same probability, it will be further displaced.

The behaviour in figure 1 is therefore explained by relation (\ref{bcp}).
Note that in figure 1(b) the value of $\mathcal{P}(E_{0}=0)$, which of course
has to be zero, is a measure for the poor quality of the zeroth order
Gordon-Volkov perturbation theory, at least in this low intensity regime.
Finally it is worth to note that the crossover which takes place for the
curves of $\alpha =1.5$ and $\alpha =2$ indicates that in fact (\ref{bcp})
is not exact and the pulse length has to be scaled also. It is not an
indication that the higher order terms need to be taken into account, since,
as we discussed in subsection 2.3,  scale invariance holds order by order in
perturbation theory.

\section{Existence criteria for stabilization}

As a consequence of (\ref{invP}) it is elementary to derive a differential
equation which has to be satisfied by the ionization probability 
\begin{equation}
\lambda \frac{d\mathcal{P}}{d\lambda }=2\tau \frac{\partial \mathcal{P}}{%
\partial \tau }-\alpha \frac{\partial \mathcal{P}}{\partial \alpha }-3E_{0}%
\frac{\partial \mathcal{P}}{\partial E_{0}}+\lambda \frac{\partial \mathcal{P%
}}{\partial \lambda }\,\,.  \label{diffe}
\end{equation}
As an example one may easily convince oneself that (\ref{dd}) indeed
satisfies (\ref{diffe}).

One way to speak of stabilization is when the ionization probability as a
function of the field amplitude satisfies 
\begin{equation}
\frac{\partial \mathcal{P}}{\partial E_{0}}\leq 0  \label{stab}
\end{equation}
for $E_{0}\in [0,\infty )$ on a finite interval. Noting now that the 
transformation of the length scale is a symmetry for the ionization
probability, i.e. relation (\ref{ion}), we have $\partial \mathcal{P}%
/\partial \lambda =$ $d\mathcal{P}/d\lambda =0$. Then, according to the
differential equation (\ref{diffe}), the criterium (\ref{stab}) for the
existence of stabilization may  be written alternatively as 
\begin{equation}
2\tau \frac{\partial \mathcal{P}}{\partial \tau }\leq \alpha \frac{\partial 
\mathcal{P}}{\partial \alpha }\,\,.  \label{crit}
\end{equation}
Once again it will be instructive to verify this statement for an explicit
example. We believe that hitherto no analytical expression for the
ionization probability is known which obeys the strict inequality in (\ref
{stab}). However, it was shown \cite{FKS2,FFS2} that in the extreme
intensity limit $E_{0}\rightarrow \infty $ the equal sign holds. In
particular when $b(\tau )=c(\tau )=0$ one obtains non-trivial expressions in
this case. Taking for instance the potential to be the $\delta $-potential
in three dimensions, the ionization probability of the only bound state was
computed to \cite{FKS2} 
\begin{equation}
\mathcal{P}\left( \varphi \right) =1-\frac{1}{\pi }\left| U\left( \frac{3}{2}%
,\frac{1}{2};\frac{i\tau \alpha ^{2}}{2}\right) \right| ^{2}\,\,\,\,\,,
\label{ex2}
\end{equation}
with $U$ being the confluent hypergeometric function. Obviously (\ref{ex2})
satisfies the criterium (\ref{crit}) for the equal sign.

It is interesting to note that for potentials which do not possess an energy
scale, like the ones mentioned after (\ref{VE}), relation (\ref{crit})
reduces to $\partial \mathcal{P}/\partial \tau \leq 0$ for $\tau \in
[0,\infty )$ on a finite interval. This means that for increasing pulse
length the ionization probability should decrease, which is as
counterintuitive as the statement (\ref{stab}).

\section{Conclusions}

We have shown that transforming the length scale corresponds to a symmetry
in the ionization probability $\mathcal{P}\left( \varphi \right) $. We
demonstrated that this symmetry property may be used as a consistency check
in various approximation methods in numerical or analytical form. One should
also note that every numerical code which fully solves the Schr\"{o}dinger
equation can be tested for consistency by appropriately scaling all
variables. Moreover one can employ the scale invariance to avoid certain
problems which sometimes  plague numerical calculations as for instance the
occurrence of very small numbers near machine precision or of very large
numbers. By re-scaling all parameters one might be able to avoid such
difficulties and still describe exactly the same physical situation.

We have further shown, in section 2.6, that certain types of behaviour for
which one has very often intuitive physical explanations may be confirmed by
means of scaling arguments.

We like to stress that none of the above considerations is restricted to a
particular intensity regime of the pulse in comparison with the potential
and they hold for low as well as ultra high intensities, although the latter
regime is of course currently of more interest. The above considerations may
of course be carried out also for other quantities of interest like
ionization rates $\mathcal{I}$, harmonic spectra etc. It follows for
instance immediately that the ionization rate has to scale inverse to the
time, i.e. $\mathcal{I}\rightarrow \mathcal{I=}\lambda ^{-\eta _{t}}\mathcal{%
I}$. Fermi's golden rule scales for instance in this way.

As an outlook one should keep in mind that like in numerous other situations
the physics becomes more interesting when the symmetry is broken. For
instance for the two-dimensional delta potential we noted already that there
is a priori no energy scale available. However, these potentials suffer from
ultraviolet divergencies at the origin which have to be renormalised.
Through this procedure one then introduces an additional scale, which is a
situation reminiscent of relativistic quantum field theory. Another
interesting situation arises when we have more than one intrinsical physical
scale in our system. In many situations one scale is dominating the other
and the problem is reducible to one with only one parameter. However, there
might intriguing situations in which the scales combine in an arbitrary
complicated manner as for instance in a statistical physics problem where we
have a microscopic length scale which specifies the typical distance between
fluctuating magnetic degrees of freedom and the correlation length.

\textbf{Acknowledgment:} A.F. and R.S. are grateful to the Deutsche
Forschungsgemeinschaft (Sfb288) for partial support. We would like to thank
M. D\"{o}rr for bringing the existence of the second reference in \cite{Lamb}
to our attention.

{\small \setlength{\baselineskip}{12pt} }

\end{document}